# Tight-Binding Device Modeling of 2-D Topological Insulator Field-Effect Transistors With Gate-Induced Phase Transition

Yungyeong Park, Yosep Park, Hyeonseok Choi, Subeen Lim, Dongwook Kim, and Yeonghun Lee, *Member, IEEE*

*Abstract*—Topological insulator field-effect transistors (TIFETs) built on 2-D quantum spin Hall insulators are being considered as advanced logic transistors due to their potentially superior performance originating from the dissipationless edge transport. This paper presents a device modeling based on the tight-binding model and the nonequilibrium Green's function formalism to simulate the current-voltage characteristics of the TIFETs. We then use the device simulator to demonstrate the effect of channel length on device performance. The device modeling will not only enable a direct estimation of TIFET performance but also shed light on the nontraditional switching operation via the topological phase transition.

## I. Introduction

The continuous advancement of semiconductor technology has led to the exploration of new materials and device concepts, such as 2-D field-effect transistors (FETs), tunnel FETs, negative capacitance FETs[1], [2]. 2D quantum spin Hall (QSH) topological insulators[3], [4] undergo exotic phase transition, where the topologically nontrivial QSH phase is transformed to the trivial normal insulator phase through the broken inversion symmetry. Unlike the trivial insulators, the 2D QSH insulators possess conducting edge states protected from backscattering by time-reversal symmetry, which provides a route to the realization of low-power devices. The phase transition due to the inversion symmetry breaking has been experimentally observed in germanene[5], $Na_3Bi$[6] and black phosphorus[7] through the measurement of the band structure and band gap modulation under different electric fields. Accordingly, the topological phase transition can be exploited for a switching operation manipulated by electrical gating as in conventional metal-oxide-semiconductor field-effect transistors (MOSFETs)[8], [9], [10], [11], [12], [13]. In addition to the straightforward gate-induced phase transition, various types of switching mechanisms—gate-modulated backscattering with intentional imperfections[14] and gate-induced additional edge states leading to conducting channels in HgTe nanoribbons[15]—have been proposed. In this sense, the topological insulator field-effect transistors (TIFETs) made of the 2D QSH insulators have emerged as an alternative device that can overcome the fundamental limits of device performance. Furthermore, the TIFETs could overcome the Boltzmann limit of the subthreshold swing, 60 mV/decade at room temperature[16], [17].

Although the TIFETs attributed to the topological phase transition may offer an unprecedented opportunity for innovation of the semiconductor technology, the integration of topological insulators into the FET framework poses significant challenges and requires a comprehensive understanding of the unique device operation. This is where device modeling becomes critical, serving as an essential tool for understanding the underlying physics of TIFETs, predicting device performance, and guiding experiments. The paper has two objectives: first, to develop a comprehensive device model that encapsulates essential device physics for the nontrivial channel materials, taking into account gate dielectric properties and channel length effects, which have previously been overlooked despite dealing with device simulations[16], [17]; and second, to utilize this model to predict the performance of TIFETs, thereby providing valuable insights for future device design and fabrication based on topological insulators.

This paper is organized as follows. In the following section, we provide a detailed description of our device modeling based on the tight-binding (TB) model and the nonequilibrium Green's function (NEGF) formalism[18], [19]. We then present how to determine the material and device parameters and demonstrate an application of the device modeling to the TIFETs with stanene[13] as the channel material, where discussing the channel length dependence of the TIFETs. Finally, we conclude this paper with potential avenues for future research.

This paragraph of the first footnote will contain the date on which you submitted your paper for review. This work was supported by Incheon National University Research Grant in 2022.

Y. Park, Y. Park, H. Choi, S. Lim, and Y. Lee are with the Department of Electronics Engineering, Incheon National University, Incheon 22012, Republic of Korea, and Y. Lee is also with the Research Institute for Engineering and Technology, Incheon National University, Incheon 22012, Republic of Korea (e-mail: y.lee@inu.ac.kr).

D. Kim is with the Department of Materials Science and Engineering, University of Utah, Salt Lake City, Utah 84112, USA.

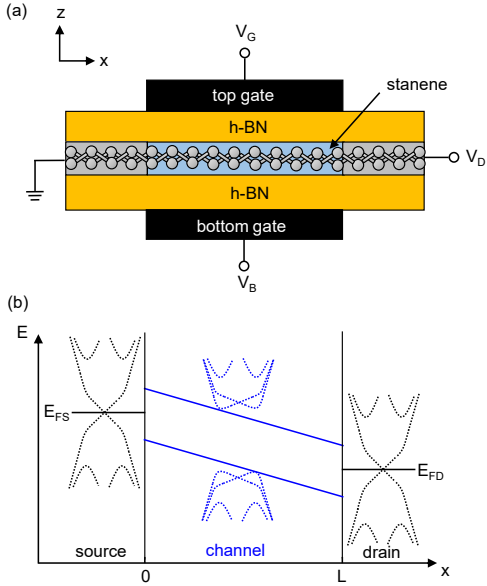

Fig. 1 Device geometry and band diagram. (a) Side view of the dual-gate device structure. The source, channel, and drain material is a zigzag stanene nanoribbon, and the gate dielectrics are hexagonal boron nitride (h-BN). (b) Schematic band diagram in real space along the channel direction (solid lines) superimposed with $E$-$k$ band structures in $k$-space (dotted lines) for the off state. The $E$-$k$ band structures assume a periodic atomic potential along with a negligible external potential gradient along the $x$-direction.

## II. Tight-Binding Device Modeling

We implement a device simulator to investigate the current-voltage characteristics of the TIFETs based on the TB model and the NEGF formalism[18], [19] using Kwant, a Python package for numerical quantum transport calculations[20]. Fig. 1(a) shows a dual-gate device structure adopted for the TIFETs. The top and bottom gates are biased by the top-gate voltage $V_\text{G}$ and the bottom-gate voltage $V_\text{B}$, respectively. The top and bottom gates are required to be biased with different voltages to break the inversion symmetry of the 2D QSH insulator. The channel potential profile is assumed to change linearly along the channel between the source and drain as depicted in Fig. 1(b). The difference between the source Fermi level $E_\text{FS}$ and the drain Fermi level $E_\text{FD}$ is given by $E_\text{FS} = E_\text{FD} + eV_\text{D}$, where $e$ is the elementary charge, and $V_\text{D}$ is the drain voltage. The channel potential energy is then expressed as $eV_\text{D}x/L$, where $L$ is the channel length.

To represent the band structure of stanene in terms of the TB model, we employ the Kane-Mele model for the QSH insulators with hexagonal lattice[3], [4]:

$$H = -t \sum_{\langle i,j\rangle\alpha} c_{i\alpha}^\dagger c_{j\alpha} + i\lambda_\text{so} \sum_{\langle\langle i,j\rangle\rangle\alpha\beta} v_{ij} c_{i\alpha}^\dagger s_{\alpha\beta}^z c_{j\beta}$$
$$+\lambda_\text{v} \sum_{i\alpha} \epsilon_i c_{i\alpha}^\dagger c_{i\alpha} + i\lambda_\text{R} \sum_{\langle i,j\rangle\alpha\beta} c_{i\alpha}^\dagger (s_{\alpha\beta} \times \hat{d}_{ij})_z c_{j\beta}, \quad (1)$$

where $c_{i\alpha}^\dagger$ ($c_{i\alpha}$) is the second-quantized creation (annihilation) operator with spin $\alpha \in \{\uparrow, \downarrow\}$ at site $i$. The first term in (1) is the nearest-neighbor hopping term from site $j$ to $i$ with a hopping strength $t$. The second term is the spin-orbit coupling (SOC) term, which is described by next-nearest-neighbor hoppings with a strength $\lambda_\text{so}$, where $v_{ij} = +1(-1)$ if the second-neighbor hopping is counterclockwise (clockwise) with respect to the positive $z$-axis. Here, $s_{\alpha\beta}^z$ are the matrix elements of the Pauli matrix $s^z$ with spins $\alpha$ and $\beta$ at sites $i$ and $j$. The third term is the staggered potential term, which break the inversion symmetry, where $\epsilon_i = +1(-1)$ for sublattice A (B). The vertical electric field $E_z$ induced by the two different gate voltages brings about the inversion symmetry breaking, meaning that the strength $\lambda_\text{v}$ is likely to be proportional to $E_z$, i.e., $\lambda_\text{v} = \alpha_\text{v} e E_z$ with a material specific parameter $\alpha_\text{v}$[16]. The last term is the Rashba term, where $s_{\alpha\beta}$ are the matrix elements of the Pauli vector $s$ with spins $\alpha$ and $\beta$ at sites $i$ and $j$, and $\hat{d}_{ij}$ are the nearest-neighbor bond unit vectors from site $j$ to $i$, $\hat{d}_{ij} = d_{ij}/|d_{ij}|$. Since the Rashba term is another source of the inversion symmetry breaking due to the our-of-plane electric field $E_z$, the strength $\lambda_\text{R}$ is also given by $\lambda_\text{R} = \alpha_\text{R} e E_z$ with a material specific parameter $\alpha_\text{R}$[16]. For the device simulation, two things need to be considered further: (1) an additional onsite energy term $\sum_{i\alpha} h_i c_{i\alpha}^\dagger c_{i\alpha}$ with $h_i = eV_\text{D}x_i/L$ to account for the linearly varying channel potential, where $x_i$ is the position of site $i$ from the source end, and (2) $\lambda_\text{v} = \lambda_\text{R} = 0$ for the source and drain to keep them in the QSH phase while the channel undergoes the topological phase transition. The schematic band structures shown in Fig. 1(b) represent the distinct phases of the source, channel, and drain for the off state, where the source, channel, and drain exhibit the QSH, trivial, and QSH phases, respectively.

Given the Hamiltonian, the NEGF method[18], [19] allows us to deal with a quantum transport phenomenon to simulate device operation. According to the Landauer-Büttiker formula[21], a drain current flowing between the source and drain leads is

$$I_\text{D} = \frac{e}{2\pi\hbar} \int T(E)[f_\text{S}(E) - f_\text{D}(E)] dE, \quad (2)$$

where $E$ is the energy, $\hbar$ is the reduced Planck's constant, $f_\text{S}(E)$ and $f_\text{D}(E)$ are the Fermi-Dirac distribution functions at the source and drain, respectively, the transmission coefficient is calculated as $T(E) = \text{Tr}\left[\Gamma_\text{S}(E) G^\text{R}(E) \Gamma_\text{D}(E) G^{\text{R}\dagger}(E)\right]$ in terms of the NEGF formalism. Here, $G^\text{R}$ is the retarded Green's function, and $\Gamma_\text{S}(E) = i[\Sigma_\text{S}(E) - \Sigma_\text{S}^\dagger(E)]$ and $\Gamma_\text{D}(E) = i[\Sigma_\text{D}(E) - \Sigma_\text{D}^\dagger(E)]$ are the broadening matrices corresponding to the source and drain leads, respectively. The retarded Green's function is

$$G^\text{R}(E) = [EI - H - \Sigma_\text{S}(E) - \Sigma_\text{D}(E)]^{-1}, \quad (3)$$

where $\Sigma_\text{S}$ and $\Sigma_\text{D}$ are the self-energies due to the two leads, and $I$ is the identity matrix.

To find the TB parameters of stanene, we carry out the first-principles electronic structure calculations based on the density functional theory (DFT)[22], [23] using Open source package for Material eXplorer (OpenMX)[24], [25], [26], [27]. We adopt the generalized gradient approximation (GGA) in the form of Perdew-Burke-Enzerhof (PBE) functional[28], norm-conserving pseudopotentials[29], and optimized pseudoatomic basis functions[24]. The real-space mesh cutoff energy is set to 220 Ry, and a $k$-point grid of $8 \times 8 \times 1$ is adopted for the Brillouin zone sampling. Noncollinear spin calculations with SOC are performed because QSH is in tandem with the SOC as included in ((1).

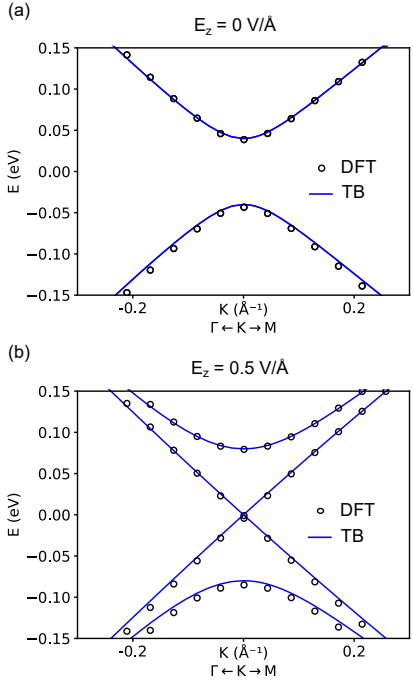

Fig. 2 Band structures of the bulk stanene around the K point to extract the TB parameters at (a) $E_z = 0$ V/Å and (b) $E_z = 0.5$ V/Å, which is the critical electric field that triggers the gap closing, at the moment of the topological phase transition. Open circles and solid lines indicate bands calculated using DFT and TB, respectively.

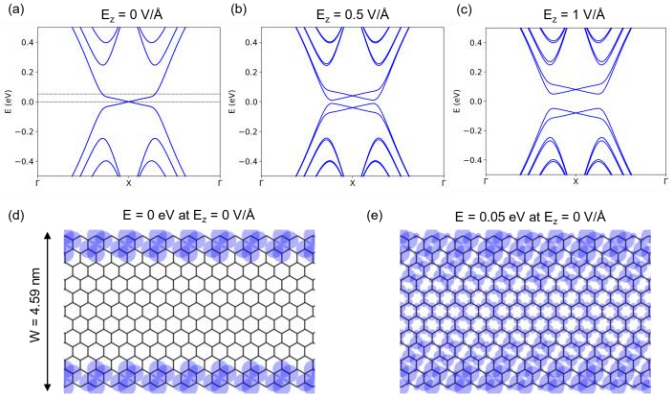

Fig. 3 Band structures of zigzag stanene nanoribbons with a width $W$ of 4.59 nm at different values of $E_z$: (a) $E_z = 0$ V/Å for the on state, and (b) $E_z = 0.5$ V/Å at threshold, and (c) $E_z = 1$ V/Å for the off state. The dotted lines in the nanoribbon band at $E_z = 0$ V/Å indicate the energy levels selected for the wave function calculation. The absolute squares of the wave functions corresponding to the states with (d) $E = 0$ eV and (e) $E = 0.05$ eV are depicted on the hexagonal lattice. The size of the circles at each site indicates the magnitude of the absolute squares in arbitrary units.

### III. Results and Discussion

The first step is to determine the parameters for the electronic structure calculations and the device simulations. Fig. 2 displays the results of fitting the TB band structures to the DFT band structures of the bulk stanene with the hexagonal lattice. The parameters $t$ and $\lambda_{so}$ are extracted in the absence of an electric field, followed by the determination of $\alpha_v$ and $\alpha_R$ through fitting at the critical electric field. The resulting values are as follows: $t = 0.7$ eV, $\lambda_{so} = 40/(3\sqrt{3})$ meV, and $\alpha_v = 0.08$ Å, and $\alpha_R = 0$ Å. The value of $\alpha_R$ being zero aligns with the negligible effect of the Rashba term in stanene[30]. For device simulation, we need additional parameters. In the device simulation, the monolayer stanene has a lattice constant of 4.68 Å, a thickness $t_s$ of 3.3 Å[31], and a relative permittivity $\varepsilon_s$ of 1.6[32]. The gate dielectrics are h-BN with a thickness $t_i$ of 3.17 Å and a relative permittivity $\varepsilon_i$ of 3.29[33]. Also, the width of nanoribbons is set to 4.59 nm, where the finite size effect is suppressed to exhibit the topological edge states[34]. Once the geometry and dielectric parameters have been set, the $E_z$ in the stanene layer is determined by taking into account series capacitors composed of the h-BN, stanene, and h-BN layers, resulting in $E_z = (V_G − V_B)/(2t_i\varepsilon_s/\varepsilon_i + t_s)$. The temperature is set to 300 K.

Fig. 3(a) shows that the QSH phase in the absence of $E_z$ exhibits the topologically protected edge states at the band crossing point around the zero energy. In this QSH phase, the backscattering is forbidden by the robust edge states in $E = 0$ eV, although the scattering between edge and bulk states exists away as the states deviate from the crossing point (Fig. 3(d) and (e)). If we break the inversion symmetry by applying a large $E_z$, the topological phase transition occurs, resulting in opening a gap (Fig. 3(c)). In the trivial phase, although the spin-momentum locked states remain near the gap, the states will suffer from backscattering due to the interaction with bulk states that emerge in the same energy levels.

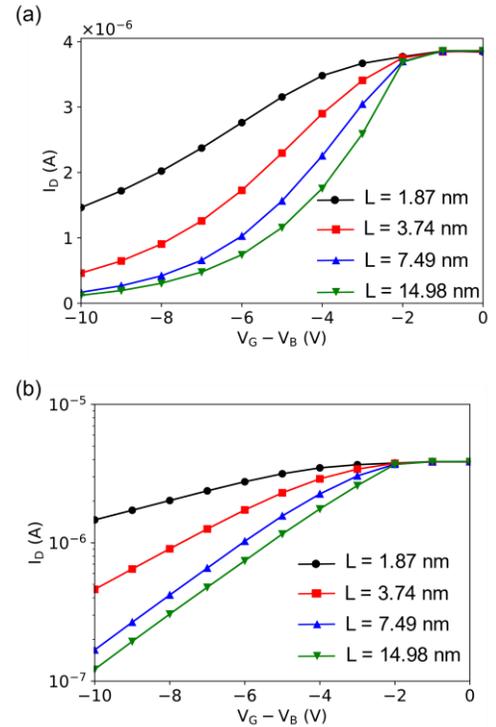

Fig. 4 Current-voltage characteristics, drain current $I_D$ versus $V_G − V_B$, in (a) linear and (b) semi-logarithmic plots for different channel lengths with $W = 4.59$ nm at $V_D = 0.05$ V.

With these material and device parameters, we perform the device simulations of stanene TIFETs with different channel lengths (Fig. 4). The $I_D$-$V_G$ curves represent a rather trivial switching operation, which is in fact desirable to replace the MOSFETs with the TIFETs, taking advantage of the matured logic semiconductor technology without much effort in terms of the current-voltage characteristics. Furthermore, a threshold voltage of the TIFETs can be adjusted by changing $V_B$, straightforwardly. Fig. 4 also shows that the on-state currents at $V_G - V_B = 0$ V does not depend on the channel length. This is not surprising because the quantum transport simulation assumes ballistic transport, which is however unrealistic for conventional MOSFETs owing to the unavoidable multiple scattering sources, such as impurity scattering and phonon scattering[35]. On the other hand, the ballistic transport is likely to take place in the TIFETs due to the forbidden backscattering of the spin-momentum locked edge states[11]. Therefore, the performance obtained in this modeling is close to the actual performance of the TIFETs.

finite conductance is observed near $E = 0$ eV for the short-channel device, while that for the long-channel device is almost zero. Along with the strong length dependence, the nonzero conductance reveals that the tunneling from the QSH source to the QSH drain through the barrier formed by the topologically trivial channel is responsible for the off leakage for the short-channel TIFET. (Fig. 5(c)) By contrast, for a long-channel devices, tunneling is prohibited due to the inability to transport carriers through the barriers of conventional channels. (Fig. 5(d))

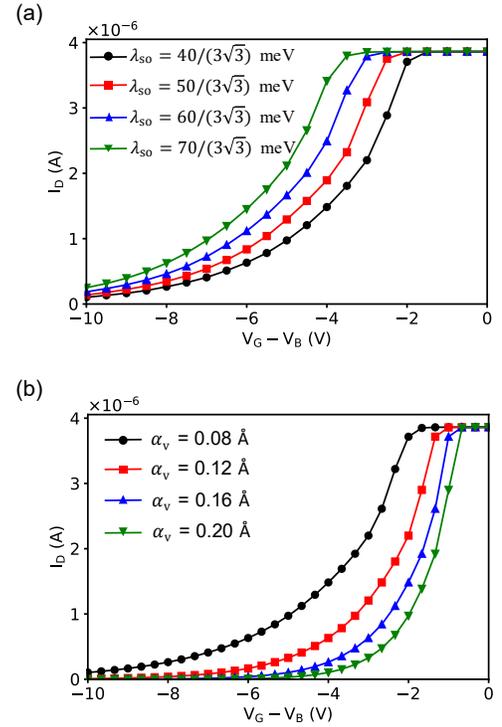

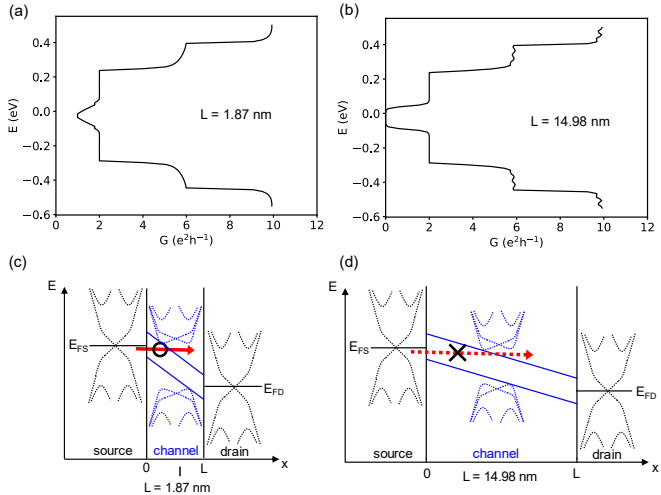

Fig. 5 Off-state conductance as a function of $E$ with (a) $L = 1.87$ nm and (b) $L = 14.98$ nm and also with $W = 4.59$ nm at $V_D = 0.05$ V and $V_G - V_B = -6.4$ V. This gate voltage produces $E_z = 1$ V/Å. The corresponding band diagrams illustrate the presence or absence of tunneling in (c) the short-channel and (d) the long-channel devices.

Fig. 3(c) shows that a large $E_z$ causes the channel stanene to open a band gap as a topological phase transition, resulting in a carrier deficiency in a trivial phase, i.e., the off state. Considering the mechanism of reaching the off state, the length dependence is not expected. Notwithstanding, the off-state current is observed to be highly dependent on the channel length. It is clearly shown that this unexpected length dependence of the off-state current is prominent for a very short channel. This length dependence brings us to the analogy of the tunneling leakage for the ultrashort-channel MOSFET[35]. Since the tunneling component can be seen directly from the conductance, we plot the off-state conductance versus energy for the shortest and the most extended devices in this work (Fig. 5(a) and (b)). The overall shape of the conductance plot for different channel lengths is similar to each other. However, a

Fig. 6 Current-voltage characteristics, drain current $I_D$ versus $V_G - V_B$, for different (a) $\lambda_{so}$ strengths and (b) $\alpha_v$ strengths with $W = 4.59$ nm and $L = 28.08$ nm at $V_D = 0.05$ V.

To improve the switching operation by reducing the wide voltage sweep range shown in Fig. 4, a device simulation was conducted with varying TB parameters. Fig. 6 depicts the $I_D$-$V_G$ curves for different $\lambda_{so}$ and $\alpha_v$, while maintaining the other stanene and device parameters at their respective fixed values. In Fig. 6(a), the device switches at a smaller gate voltage as the SOC strength decreases. This is because a small SOC reduces the band gap of the QSH phase, resulting in a reduction of the gate voltage required for the phase transition. Also, Fig. 6(b) shows that the reduction of the gate voltage requited for the phase transition is accompanied with a large $\alpha_v$. This is due to the fact that when a perpendicular electric field is applied, the increased staggered potential ($\lambda_v = \alpha_v e E_z$) breaks the spin-valley degeneracy and reduces the gap rapidly, which leads to a spin-polarized gapless phase and the phase transition, eventually. This tendency presents potential routes for low-voltage operation and performance optimization of TIFETs.

A further discussion can be made based on the results of the TIFET modeling. The edge characteristics of 2D topological insulators are unlikely to be fully reproduced by solely relying on TB parameters extracted from the bulk band structure. It is necessary to analyze edge transport of TIFETs at a microscopic level, taking into account various edge characteristics, such as edge relaxation, passivation, and types of nanoribbons (e.g., armchair and zigzag). Also, having considered the large gate voltage sweep range reaching around 10 V required for switching the device with the atomically thin gate dielectrics, the stanene is unlikely to be a suitable channel material for TIFETs. Of course, this does not mean all TIFETs are impractical. There are various 2D topological insulators with different parameters[36] that could resolve this issue and further improve the TIFET. For example, 1T'-$MoS_2$ undergoes the topological phase transition at $E_z = 0.142$ V/Å—a much smaller critical electric field than that of the stanene—by controlling the Rashba term[11], which may allow the device to operate with a steep slope.

## IV. Conclusion

The device modeling of the TIFETs was performed with the TB model and the NEGF formalism. In doing so, we presented a detailed procedure spanning from the mathematical formalisms to the parameter extraction. Apart from that, the simulation results for the devices with varying channel lengths disclosed that a substantially long channel is necessary to diminish the tunneling portion of the off-leakage current as in the ultrashort-channel MOSFETs. The simulation results conducted using various material parameters indicated promising avenues for enhancing TIFET performance. Nevertheless, it is evident that further investigation is required to achieve steep-slope and low-power devices. The device modeling allowed us to study how the exotic topological phase transition manipulates the device operation and can be extended to the investigation of other intriguing features of the TIFETs, including geometry effects, junction engineering, nonuniform gate control over the channel, and so on. At the end of the day, this work will contribute to the fundamental understanding of the interplay between topological physics and semiconductor device engineering.

## References

[1] S. Salahuddin, K. Ni, and S. Datta, "The era of hyper-scaling in electronics," *Nature Electronics*, vol. 1, no. 8, pp. 442–450, Aug. 2018, doi: 10.1038/s41928-018-0117-x.
[2] S. Das *et al.*, "Transistors based on two-dimensional materials for future integrated circuits," *Nat Electron*, vol. 4, no. 11, pp. 786–799, Nov. 2021, doi: 10.1038/s41928-021-00670-1.
[3] C. L. Kane and E. J. Mele, "Z 2 Topological Order and the Quantum Spin Hall Effect," *Phys. Rev. Lett.*, vol. 95, no. 14, p. 146802, Sep. 2005, doi: 10.1103/PhysRevLett.95.146802.
[4] C. L. Kane and E. J. Mele, "Quantum Spin Hall Effect in Graphene," *Phys. Rev. Lett.*, vol. 95, no. 22, p. 226801, Nov. 2005, doi: 10.1103/PhysRevLett.95.226801.
[5] P. Bampoulis *et al.*, "Quantum Spin Hall States and Topological Phase Transition in Germanene," *Phys. Rev. Lett.*, vol. 130, no. 19, p. 196401, May 2023, doi: 10.1103/PhysRevLett.130.196401.
[6] J. L. Collins *et al.*, "Electric-field-tuned topological phase transition in ultrathin Na3Bi," *Nature*, vol. 564, no. 7736, pp. 390–394, Dec. 2018, doi: 10.1038/s41586-018-0788-5.
[7] J. Kim *et al.*, "Observation of tunable band gap and anisotropic Dirac semimetal state in black phosphorus," *Science*, vol. 349, no. 6249, pp. 723–726, Aug. 2015, doi: 10.1126/science.aaa6486.
[8] L. A. Wray, "Topological transistor," *Nature Phys*, vol. 8, no. 10, pp. 705–706, Oct. 2012, doi: 10.1038/nphys2410.
[9] M. Ezawa, "Quantized conductance and field-effect topological quantum transistor in silicene nanoribbons," *Applied Physics Letters*, vol. 102, no. 17, p. 172103, Apr. 2013, doi: 10.1063/1.4803010.
[10] J. Liu, T. H. Hsieh, P. Wei, W. Duan, J. Moodera, and L. Fu, "Spin-filtered edge states with an electrically tunable gap in a two-dimensional topological crystalline insulator," *Nature Mater*, vol. 13, no. 2, pp. 178–183, Feb. 2014, doi: 10.1038/nmat3828.
[11] X. Qian, J. Liu, L. Fu, and J. Li, "Quantum spin Hall effect in two-dimensional transition metal dichalcogenides," *Science*, vol. 346, no. 6215, pp. 1344–1347, Dec. 2014, doi: 10.1126/science.1256815.
[12] Q. Liu, X. Zhang, L. B. Abdalla, A. Fazzio, and A. Zunger, "Switching a Normal Insulator into a Topological Insulator via Electric Field with Application to Phosphorene," *Nano Lett.*, vol. 15, no. 2, pp. 1222–1228, Feb. 2015, doi: 10.1021/nl5043769.
[13] A. Molle, J. Goldberger, M. Houssa, Y. Xu, S.-C. Zhang, and D. Akinwande, "Buckled two-dimensional Xene sheets," *Nature Mater*, vol. 16, no. 2, pp. 163–169, Feb. 2017, doi: 10.1038/nmat4802.
[14] W. G. Vandenberghe and M. V. Fischetti, "Imperfect two-dimensional topological insulator field-effect transistors," *Nat Commun*, vol. 8, no. 1, p. 14184, Jan. 2017, doi: 10.1038/ncomms14184.
[15] H.-H. Fu, J.-H. Gao, and K.-L. Yao, "Topological field-effect quantum transistors in HgTe nanoribbons," *Nanotechnology*, vol. 25, no. 22, p. 225201, Jun. 2014, doi: 10.1088/0957-4484/25/22/225201.
[16] M. Nadeem, I. Di Bernardo, X. Wang, M. S. Fuhrer, and D. Culcer, "Overcoming Boltzmann's Tyranny in a Transistor via the Topological Quantum Field Effect," *Nano Lett.*, vol. 21, no. 7, pp. 3155–3161, Apr. 2021, doi: 10.1021/acs.nanolett.1c00378.
[17] S. Banerjee, K. Jana, A. Basak, M. S. Fuhrer, D. Culcer, and B. Muralidharan, "Robust Subthermionic Topological Transistor Action via Antiferromagnetic Exchange," *Phys. Rev. Applied*, vol. 18, no. 5, p. 054088, Nov. 2022, doi: 10.1103/PhysRevApplied.18.054088.
[18] L. V. Keldysh, "Diagram Technique for Nonequilibrium Processes," *JETP*, vol. 20, p. 1018, 1965.
[19] S. Datta, "Electrical resistance: an atomistic view," *Nanotechnology*, vol. 15, no. 7, pp. S433–S451, Jul. 2004, doi: 10.1088/0957-4484/15/7/051.
[20] C. W. Groth, M. Wimmer, A. R. Akhmerov, and X. Waintal, "Kwant: a software package for quantum transport," *New J. Phys.*, vol. 16, no. 6, p. 063065, Jun. 2014, doi: 10.1088/1367-2630/16/6/063065.
[21] R. Landauer, "Spatial Variation of Currents and Fields Due to Localized Scatterers in Metallic Conduction," *IBM J. Res. & Dev.*, vol. 1, no. 3, pp. 223–231, Jul. 1957, doi: 10.1147/rd.13.0223.
[22] P. Hohenberg and W. Kohn, "Inhomogeneous Electron Gas," *Phys. Rev.*, vol. 136, no. 3B, pp. B864–B871, Nov. 1964, doi: 10.1103/PhysRev.136.B864.
[23] W. Kohn and L. J. Sham, "Self-Consistent Equations Including Exchange and Correlation Effects," *Phys. Rev.*, vol. 140, no. 4A, pp. A1133–A1138, Nov. 1965, doi: 10.1103/PhysRev.140.A1133.
[24] T. Ozaki, "Variationally optimized atomic orbitals for large-scale electronic structures," *Phys. Rev. B*, vol. 67, no. 15, p. 155108, Apr. 2003, doi: 10.1103/PhysRevB.67.155108.
[25] T. Ozaki and H. Kino, "Numerical atomic basis orbitals from H to Kr," *Phys. Rev. B*, vol. 69, no. 19, p. 195113, May 2004, doi: 10.1103/PhysRevB.69.195113.
[26] T. Ozaki and H. Kino, "Efficient projector expansion for the *ab initio* LCAO method," *Phys. Rev. B*, vol. 72, no. 4, p. 045121, Jul. 2005, doi: 10.1103/PhysRevB.72.045121.
[27] K. Lejaeghere *et al.*, "Reproducibility in density functional theory calculations of solids," *Science*, vol. 351, no. 6280, p. aad3000, Mar. 2016, doi: 10.1126/science.aad3000.
[28] J. P. Perdew, K. Burke, and M. Ernzerhof, "Generalized Gradient Approximation Made Simple," *Phys. Rev. Lett.*, vol. 77, no. 18, pp. 3865–3868, Oct. 1996, doi: 10.1103/PhysRevLett.77.3865.
[29] I. Morrison, D. M. Bylander, and L. Kleinman, "Nonlocal Hermitian norm-conserving Vanderbilt pseudopotential," *Phys. Rev. B*, vol. 47, no. 11, pp. 6728–6731, Mar. 1993, doi: 10.1103/PhysRevB.47.6728.
[30] K. Jana and B. Muralidharan, "Robust all-electrical topological valley filtering using monolayer 2D-Xenes," *npj 2D Mater Appl*, vol. 6, no. 1, p. 19, Mar. 2022, doi: 10.1038/s41699-022-00291-y.
[31] S. Saxena, R. P. Chaudhary, and S. Shukla, "Stanene: Atomically Thick Free-standing Layer of 2D Hexagonal Tin," *Sci Rep*, vol. 6, no. 1, p. 31073, Aug. 2016, doi: 10.1038/srep31073.
[32] R. John and B. Merlin, "Optical properties of graphene, silicene, germanene, and stanene from IR to far UV – A first principles study,"

*Journal of Physics and Chemistry of Solids*, vol. 110, pp. 307–315, Nov. 2017, doi: 10.1016/j.jpcs.2017.06.026.

[33] A. Laturia, M. L. Van de Put, and W. G. Vandenberghe, "Dielectric properties of hexagonal boron nitride and transition metal dichalcogenides: from monolayer to bulk," *npj 2D Mater Appl*, vol. 2, no. 1, p. 6, Dec. 2018, doi: 10.1038/s41699-018-0050-x.

[34] C.-H. Chung, D.-H. Lee, and S.-P. Chao, "Kane-Mele Hubbard model on a zigzag ribbon: Stability of the topological edge states and quantum phase transitions," *Phys. Rev. B*, vol. 90, no. 3, p. 035116, Jul. 2014, doi: 10.1103/PhysRevB.90.035116.

[35] S. M. Sze, K. K. Ng, and Y. Li, *Physics of semiconductor devices*, Fourth edition. Hoboken, NJ, USA: Wiley, 2021.

[36] M. S. Lodge, S. A. Yang, S. Mukherjee, and B. Weber, "Atomically Thin Quantum Spin Hall Insulators," *Advanced Materials*, vol. 33, no. 22, p. 2008029, Jun. 2021, doi: 10.1002/adma.202008029.